%% file: mhiir.tex
\documentclass[fleqn,usenatbib]{mnras}

\usepackage{amsmath}	

\usepackage{txfonts}

\usepackage[T1]{fontenc}
\usepackage{ae,aecompl}

\usepackage{hyperref}	
\hypersetup{colorlinks=true,linkcolor=blue,citecolor=blue,filecolor=blue,urlcolor=blue}
\usepackage{graphicx}	
\usepackage{amssymb}	
\usepackage{color}

\input{macros.tex}

\title[Magnetic Amplification around the First Stars]{Amplification of
  Magnetic Fields in a Primordial \hii{} Region and Supernova}

\author[Koh \& Wise]{
Daegene Koh,$^{1}$\thanks{E-mail: dkoh30@gatech.edu}
John H. Wise,$^{1}$\thanks{E-mail: jwise@gatech.edu}
\\
$^{1}$Center for Relativistic Astrophysics, Georgia Institute of Technology, 837 State Street, Atlanta, GA 30332, USA\\
}

\pubyear{2016}

\begin{document}
\label{firstpage}
\pagerange{\pageref{firstpage}--\pageref{lastpage}}
\maketitle

\begin{abstract}

  Magnetic fields permeate the Universe on all scales and play a key
  role during star formation.  We study the evolution of magnetic
  fields around a massive metal-free (Population III) star at $z \sim
  15$ during the growth of its \hii{} region and subsequent supernova
  explosion by conducting three cosmological magnetohydrodynamics
  simulations with radiation transport.  Given the theoretical
  uncertainty and weak observational constraints of magnetic fields
  in the early universe, we initialize the simulations with identical
  initial conditions only varying the seed field strength.
  We find that magnetic fields grow as $\rho^{2/3}$ during the
  gravitational collapse preceding star formation, as expected from
  ideal spherical collapse models.  Massive Population III stars can
  expel a majority of the gas from the host halo through radiative
  feedback, and we find that the magnetic fields are not amplified
  above the spherical collapse scaling relation during this phase.
  However, afterwards when its supernova remnant can radiatively cool and
  fragment, the turbulent velocity field in and around the shell
  causes the magnetic field to be significantly amplified on average
  by $\sim$100 in the shell and up to 6 orders of magnitude behind the
  reverse shock.  Within the shell, field strengths are on
  the order of a few nG at a number density of 1 cm$^{-3}$.  We show that
  this growth is primarily caused by small-scale dynamo action in the remnant.
  These strengthened fields will propagate into the first generations
  of galaxies, possibly affecting the nature of their star formation.


\end{abstract}




\section{Introduction}

Magnetic fields are everywhere in the present day universe \citep[see][for a review]{Beck1996}. Various observations reveal the presence of magnetic fields at scales ranging from planets all the way to the voids between large cosmological structures \citep{Kronberg1994, Beck1999}. Moreover, measurements of galaxies show corresponding field strengths of up to 10s of $\mu$G \citep{Beck2009}.

Such fields may originate from the amplification of primordial fields in the early universe. These primordial fields may have been generated during the electroweak and QCD phase transitions \citep{Sigl1997}.   Furthermore, \citet{Wagstaff2014} demonstrated that sufficient turbulent conditions are realized in the radiation dominated universe prior to the onset of structure formation to produce field strengths on the order of $B_0^{\mathrm{rms}} \sim (10^{-6} - 10^{-3})$ nG on scales of 0.1 - 100 pc, sufficient to explain the magnetic field strengths found in the intergalatic medium \citep[IGM;][]{Neronov2010}. Alternatively, \citet{Naoz2013} found that primordial magnetic fields are expected to be generated through the Biermann battery mechanism \citep{Biermann1950} during linear structure formation through vorticity produced by scale-dependent temperature fluctuations.  

On the other hand, the study of Population III star formation has been largely carried out without the addition of such magnetic fields. Earlier, these stars were thought to have been massive $M_\star \sim$ 100 $\mathrm{M}_{\odot}$ with suppressed fragmentation largely forming in isolation \citep{Abel2002}. However, follow up studies with longer integration times at higher densities resulted in fragmentation, suggesting that Population III binaries are possible \citep{Turk2009, Greif2011, Susa2014}.  In particular, metal-free gravitational collapses in cosmological simulations have been followed until the formation of a protostellar shock \citep{Yoshida2007}, capturing the dynamics and fragmentation of the surrounding accretion disk \citep{Greif2012}.  In the very early stages of disk fragmentation, the majority of protostars have masses $M_\star < 1~\mathrm{M}_\odot$ and some might be ejected from the central system \citep{Greif2012, Stacy2016}.  The final stellar masses are ultimately determined when the protostellar radiation quenches the accretion flow.  Most recently, \citet{Hirano2015} followed the formation and evolution of 1540 Pop III star-forming clouds, extracted from a cosmological simulation with a far-ultraviolet radiation background, with axisymmetric radiation hydrodynamic simulations. They found two distinct populations of metal-free, those formed in relative isolation versus those formed under the influence of $\rm{H}_2$-dissociating external feedback. They found an initial mass function (IMF) with two peaks at $M_\star \simeq 250~\mathrm{M}_{\odot}$ and 25 $\mathrm{M}_{\odot}$ for the former population and a single peak at $M_\star \simeq 400~\mathrm{M}_{\odot}$ for the latter population, demonstrating that metal-free star formation could indeed favor a top-heavy IMF.

As these stars begin to emit ionizing photons, they photoionize and photoheat their host halos and surrounding medium, creating a cosmological \hii{} region. The particular radiative characteristics of Pop III stars were explored by \citet{Tumlinson2000} and \citet{Schaerer2002} using evolutionary synthesis models. The latter results were then taken to study the resulting \hii{} regions in one-dimensional hydrodynamics calculations \citep{Whalen2004, Kitayama2004} showing that they span a typical radius of 1--3 kpc. Follow up three-dimensional studies with radiative transfer largely confirmed these results \citep{Alvarez2006, Abel2007}. At the end of its lifetime, the star dies in a Type II core collapse supernova for $11 \la M_\star/$$\mathrm{M}_{\odot} \la 40$ \citep{Woosley1995}, or in a pair-instability supernova for $140 \la M_\star/\mathrm{M}_{\odot} \la 260$ \citep{Heger2002}. These forms of stellar feedback were incorporated in numerical studies performed by \citet{Kitayama2005}, \citet{Greif2007}, and  \citet{Whalen2008} tracing the near complete evacuation of baryons from the host halo. In particular, \citet{Greif2007} characterized the behavior of the SN remnant in a numerical study following the four classical distinct sequential phases \citep[e.g.][]{Ostriker1988}: free expansion, Sedov-Taylor, pressure-driven snowplow, and momentum-conserving snowplow.  Mixing of heavy elements expelled from the first stars can lead to fragmentation and low-mass metal-enriched star formation in neighboring minihalos and direct halo descendants, hosting the first galaxies \citep{Wise2008, Greif2010, Smith2015}.

Both analytic and numerical studies have demonstrated the amplification of a seed magnetic field by small scale dynamos during the collapse of primordial halos \citep{King2005, Schleicher2009}. In the absence of turbulence or other dynamo action, gravitational collapse can enhance the magnetic field strength as $B \propto \rho^{2/3}$ assuming the field is frozen to the fluid. Building upon this analytical work, \citet{Sur2010} inserted a seed field of $B_{\mathrm{rms}} \sim 1$ nG into an isolated Bonnor-Ebert sphere, resulting in fields $\sim 10^{-3}$ G at a baryon density $n \sim 10^{14}~\mathrm{cm}^{-3}$. Such fields may become dynamically important in subsequent star formation by potentially reducing fragmentation of molecular clouds \citep{Clark2011}. Even without a seed field, \citet{Xu2008} showed that significant fields can be formed through the Biermann battery effects. They found a peak magnetic field strength of 1 nG at a baryon density $n \sim 10^{10}~\mathrm{cm}^{-3}$ at the center of the star forming halo at $z \simeq 18$. These fields, resulting from the Biermann term, are never strong enough to become dynamically important, but rather set a lower bound on fields that would exist during Pop III star formation.

Furthermore, \citet{Federrath2011} simulated the collapse of an isothermal Bonnor-Ebert sphere with a seed magnetic field and turbulent velocity fields showing that a minimum resolution of 32 elements per Jeans length is required to properly resolve dynamo action. As they increased the resolution up to 128 elements, they found significantly increased amplification rates with no signs of convergence. \citet{Turk2012} then performed a full numerical calculation from cosmological initial conditions demonstrating similar results. They also found that a minimal resolution of 64 elements per Jeans length is required fully capture vortical motions that can enhance magnetic fields. These results imply the need for a much more stringent resolution requirement to fully explore Pop III star formation.

Thus far, these works have all mainly focused on the generation and evolution of magnetic fields during the primordial collapse, but they all stop short of the formation of the star. In this paper, we present calculations following the evolution of magnetic fields throughout the formation, main sequence, and aftermath of a Pop III star starting from cosmological initial conditions. We follow the magnetic amplification rates as the supernova remnant expands into the surrounding medium. In the following section, we describe the specifics of the numerical simulations. In Section \ref{sec:Results}, we present the amplification of the initial background magnetic field. We then discuss the missing physics that may potential influence our results in Section \ref{sec:disc}. Finally, we summarize our results in Section \ref{sec:conc}. 

\section{Methods}
\label{sec:Methods}

\subsection{Simulation Setup}

The simulations described subsequently have all been conducted with the \enzo{} simulation code v2.4 \citep{Bryan2014}. \enzo{} is an adaptive mesh refinement (AMR) code that uses an N-body adaptive particle-mesh solver to follow dark matter dynamics. We utilize a nine-species (\hi, \hii, \hei, \heii, \heiii, $\mathrm{e^-}$, $\mathrm{H_2}$, $\mathrm{H_{2}^{+}}$, $\mathrm{H^-}$) non-equilibrium chemistry model \citep{Abel1997} using the $\mathrm{H_2}$ cooling rates from \citet{Glover2008}. To solve the ideal magneto-hydrodynamical (MHD) equations, we use the Godunov MUSCL (monotone upstream-centered schemes for conservation laws) algorithm with the Dedner hyperbolic cleaning method to enforce $\nabla \cdot B = 0$ \citep{Dedner2002, Wang2009}. We also use the Harten-Lax-van-Leer (HLL) Riemann solver with piecewise linear reconstruction for accurate shock capturing.

We initialized the simulation at $z = 150$ with a $250 h^{-1}$ comoving kpc box. The initial conditions were generated with the MUSIC initial condition generator \citep{Hahn2011} using second-order Lagrangian perturbation theory and the Planck 2013 best fit cosmological parameters \citep{PlanckCollaboration2014}: $\Omega_{\rm M} = 0.3175$, $\Omega_\Lambda=0.6825$, $\Omega_b = 0.049$, $h = 0.6711$, $\sigma_8 = 0.8344$, and $n_{\rm s}$ = 0.9624 with the symbols having their typical definitions.

First, we ran a dark matter only simulation with a $256^3$ top grid with 8 levels of adaptive mesh refinement to $z = 12$. Next, we used the Rockstar halo finder \citep{Behroozi2013a} to identify the most massive halo with a virial mass $M_{\rm vir} = 2.3 \times 10^{6} \mathrm{M}_{\odot}$ and radius $r_{\rm vir} = 316 \unit{pc}$. We then calculate the initial Lagrangian volume centered on this halo that is a sphere with a radius of $4r_{\rm vir}$.  The zoom-in initial conditions have two nested grids around this Lagrangian volume at $z=150$.  The effective dark matter mass resolution is 1.6 $\mathrm{M}_{\odot}$ in the high-resolution region, which is bounded by a cuboid with dimensions of ($72.3 \times 70.3 \times 76.2$) comoving kpc$^3$ resolved by ($296 \times 288 \times 312$) cells. We only allow the mesh to be refined in the exact Lagrangian volume of this sphere up to a maximum level of 15, corresponding to a maximal comoving spatial resolution of 0.04 pc.

The cells are flagged for refinement if one or more of the following criteria are met: (i) relative baryon overdensity of 3, (ii) relative DM overdensity of 3, and (iii) local Jeans length \citep{Truelove1997}. For the first criteria, we employ super-Lagrangian refinement, where the cells are refined more aggressively, i.e. a lower density refinement threshold, at higher levels\footnote{This feature is triggered with the \enzo{} parameter {\tt MinimumMassForRefinementLevelExponent} = --0.2 \citep[see][for more details]{Bryan2014}.}. We also require the local Jeans length to be covered by at least 64 cells in each direction in order to fully resolve the vortical motions that can amplify the magnetic field as demonstrated by \citet{Federrath2011} and \citet{Turk2012}.

Furthermore, a time-dependent Lyman-Warner optically thin radiation background modeled in \citet{Wise2012a} is utilized in the simulation, which is based on the semi-analytical model of \citet{Wise2005}.  This model considers the LW contributions of both Pop III stars and galaxies and is valid at higher redshifts ($z \ga 12$) before metal-enriched stars dominate the cosmic emissivity.  We use the functional form of the background evolution in \citet{Wise2012},
\begin{equation}
\label{eqn:lwb}
\log_{10} J_{21}(z) = A + Bz + Cz^2 + Dz^3 + Ez^4, 
\end{equation}
where (A, B, C, D, E) = (-2.567, 0.4562, -0.02680, 5.882 $\times$ 10$^{-4}$, -5.056 $\times$ 10$^{-6}$), and $J_{21}$ is the specific intensity in units of 10$^{-21}$ erg s$^{-1}$ cm$^{-2}$ Hz$^{-1}$ sr$^{-1}$.  Modulating this background, we include a prescription for radiative self-shielding taken from \citet{Wolcott-Green2011} which reduces LW flux to supress $\mathrm{H_2}$ cooling in haloes.

Each simulation was evolved until the most massive halo undergoes catastrophic cooling and collapse, and we momentarily stop the simulation at a refinement level of 15. We outputted data every 24.2 Myrs until this point. Once the halo collapsed, we then allow for star formation and feedback and wrote data every $10^{5}$ yr until the end of the simulation, 2 Myr after the supernova.  By writing data at a relatively small time interval, we are able to trace the evolution of the regions around the star and ensuing supernova.  The runs all end around $z = 14.4$. We ran the simulations on the Comet supercomputer at the San Diego Supercomputing Center using 12 nodes with 12 cores per node for each simulation. The runs took approximately 10 days each for a total computational time of 100,000 core hours. All of the analysis were performed with the analysis and visualization toolkit {\sc yt} \citep{Turk2011a}.

\subsection{Initial Magnetic Field}

We conducted a total of three runs. Each run used the same initial conditions described previously. At the start of each simulation, we seed the box with a initial uniform background field of a given field strength purely in the $z$-direction. The seed fields are given in proper magnetic field strengths that are proportional to the square of the scale factor. The only difference between the runs is the initial seed magnetic field strength. Observations of high-energy photons from blazars put the lower limit of a background field at $10^{-15}$ G \citep{Dolag2011} while the upper limit on the field strength produced by primordial phase transitions is at $10^{-20}$ G \citep{Sigl1997}. Globally, the most recent constraint from CMB measurements puts the upper limit for the comoving field strength at scale of 1 Mpc at 4.4 nG \citep{PlanckCollaboration2015}. Given the large uncertainty in the background field strength, we chose three different values. In the base case run, H2R, there is no seed magnetic field. In the runs H2R.B1 and H2R.B2, a proper seed field of $10^{-10}$ G and $10^{-14}$ G, respectively, was placed at the start. These correspond to comoving fields strengths of $4.4 \times 10^{-15}$ G and $4.4 \times 10^{-19}$ G. 

\subsection{Star Formation and Feedback}
\label{sec:starform} 

We only consider Pop III star formation in this work, and here we briefly describe the prescription for the formation and subsequent feedback mechanisms.  We represent a single Pop III star using a single star particle \citep{Abel2007, Wise2012}. A particle is formed in a cell when the following criteria have been met.
\begin{enumerate}
  \item An overdensity of $1 \times 10^6$ $(\sim6000$ $\textrm{cm}^{-3}$ at $z = 15)$.
  \item A converging gas flow ($\nabla \cdot \mathbf{v}_{\rm gas} < 0$).
  \item A molecular hydrogen fraction $f_{\mathrm{H}_2} > 5 \times 10^{-4}$.
\end{enumerate}
Given the uncertainty about the initial mass function of Pop III stars, we chose a fixed mass of 40 $\mathrm{M}_{\odot}$ as the stellar mass. Then after the formation criteria are met, an equal mass of gas is then removed from the computation grid in a sphere containing twice the stellar mass and is centered on the particle. This particle is then initialized with the mass-weighted velocity of gas contained in the sphere. Moreover, we manually limited the simulation to prevent the formation of any subsequent stars after the first star was formed to minimize the computational stress of following multiple halos since our focus was only on the most massive halo.

\begin{figure*} 
	\includegraphics[width=\textwidth]{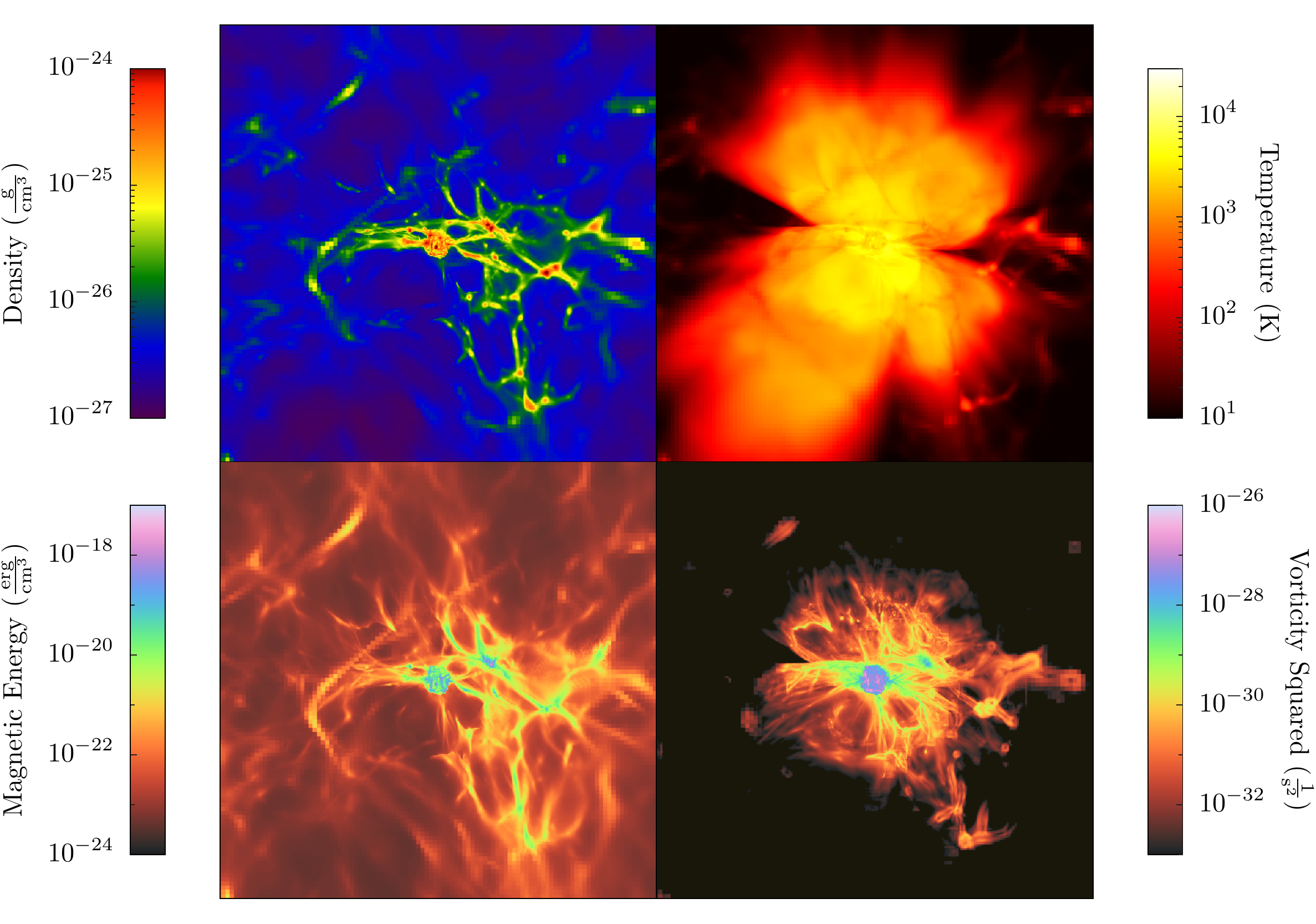}
    \caption{Mass-weighted projections of density, temperature, magnetic energy, and vorticity squared centered around the most massive halo at the end of the B1 run. Each projection has a width of 10 kpc. The \hii{} region produced by the star is most prominently displayed in the temperature plot.}
    \label{fig:proj}
\end{figure*}

After the formation, the star particle becomes a point source of H$_2$-dissociating, hydrogen- and helium-ionizing radiation.  For the dissocating radiation, we approximate the radiation intensity as a $1/r^2$ field that is centered at the star particle, providing additional intensity on top of the background (Equation \ref{eqn:lwb}).  The ionizing radiation field is evolved with adaptive ray tracing based on the HEALPix framework and is coupled self-consistently to the hydrodynamics \citep{Wise2011}. As the rays propagate outwards from the source, they are adaptively split into child rays when the solid angle associated with the parent ray $\theta = 4\pi / (12 \times 4^{L}) $, where L is the HEALpix level, is larger than 20\% the cell area. We use a discretized spectrum for the radiation with the following luminosities and photon energies: for H$_2$ dissociating radiation, $L_\gamma = 2.90 \times 10^{49}~\mathrm{s}^{-1}$; for hydrogen ionizing photons, $L_{\gamma} = 2.47 \times 10^{49}~\mathrm{s}^{-1}$ and $E_{\mathrm{ph}} = 28$ eV, which is appropriate for the near-constant $10^5$ K surface temperatures of Pop III stars; we also have helium singly and doubly ionizing radiation with lumonisities and photon energies of $L_{\gamma} = 1.32 \times 10^{49}~\mathrm{s}^{-1}$, $E_{\mathrm{ph}} = 30$ eV and $L_{\gamma} = 8.80 \times 10^{46}~\mathrm{s}^{-1}$, $E_{\mathrm{ph}} = 58$ eV, respectively \citep{Schaerer2002}.  At the end of its lifetime of 3.7 Myr, the star particle dies as a Type II supernova with a standard explosion energy of $10^{51}$ erg. The blast wave produced is modeled by injecting the thermal energy and ejecta mass into a sphere with a 5 pc radius. This injection is smoothed over the surface for numerical stability and is well resolved at initialization showing agreement with the Sedov-Taylor solution \citep{Wise2008}.

\section{Results}
\label{sec:Results}

We focus on the evolution of the magnetic field strength and morphology through the formation, main sequence, and supernova of a Pop III star, paying special attention to the amplification of primordial magnetic fields as the gas is processed by stellar radiation and the supernova. First, we visually inspect any morphological differences between the three simulations with varying initial magnetic field strengths.  We then quantify any field amplification that is caused by small dynamo actions beyond the expected compressional amplification.

\subsection{Visual inspection}

\begin{figure*} 
	\includegraphics[width=\textwidth]{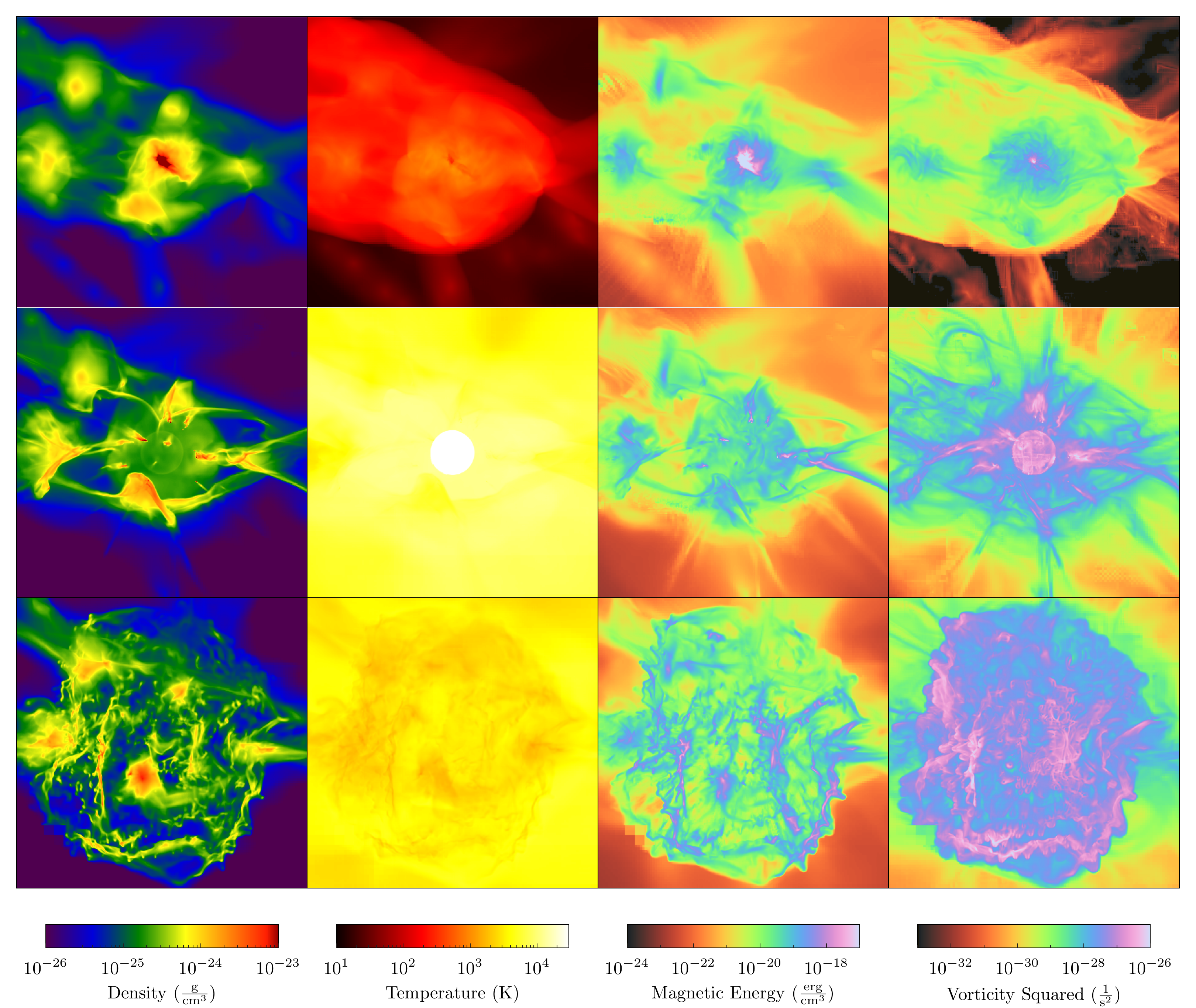}
    \caption{Density-weighted projections of density, temperature, magnetic field, and vorticity squared for the B1 run at three different times. Each projection has a field of view of 700 pc. The top panels show projections immediately following the birth of the star. The middle panels show the death of the star, and then 2 Myr after the supernova explosion at the bottom. Significant magnetic energy and vorticity is generated in the supernova remnant. The vorticity projection shows some grid artifacts as a result of the rendering which does not reflect the data.}
    \label{fig:projzoom}
\end{figure*}

\begin{figure*} 
	\includegraphics[width=\textwidth]{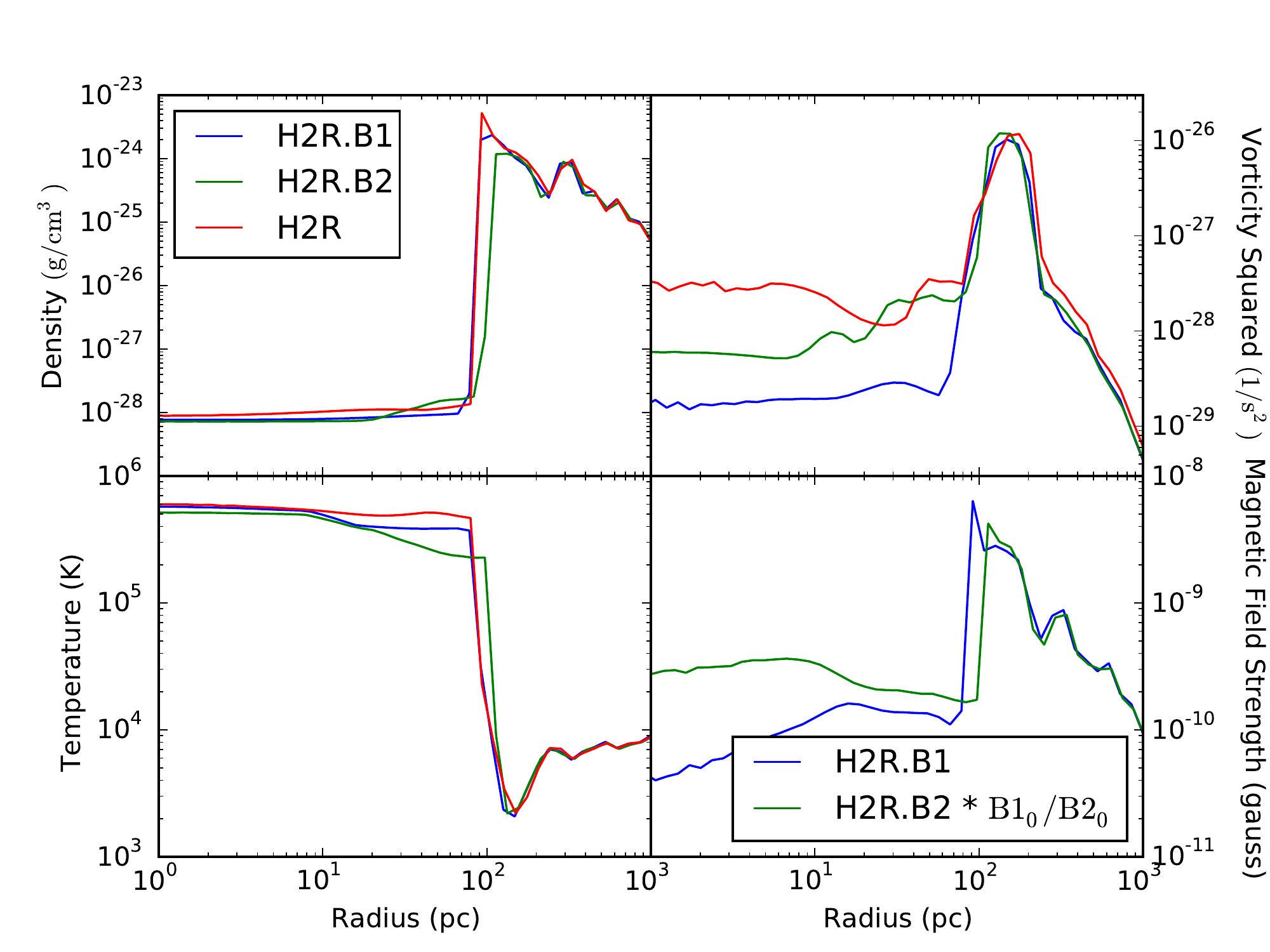}
        \caption{Radial profiles of density, temperature, magnetic field strength, and vorticity squared for all three runs at the end of the simulation 2 Myr after the supernova, centered about the star particle. In the magnetic field profiles, the field strengths in the H2R.B2 run have been scaled by $\mathrm{B}1_0 / \mathrm{B}2_0$ where $\mathrm{B}1_0, \mathrm{B}2_0$ are the initial seed values of $10^{-10}$ G and $10^{-14}$ G respectively for better comparison. }
    \label{fig:profiles}
\end{figure*}

The most massive halo has a mass of $6.0 \times 10^5$ $\mathrm{M}_{\odot}$ at the time of collapse at $z = 14.66$. A Pop III star forms near the center of the halo and begins to emit radiation heating up the entire region. Figure \ref{fig:proj} shows projections of the entire \hii{} region at the end of main sequence spanning a 10 kpc box. All three simulations have nearly identical characteristics at this time. The temperature projection clearly shows the extent of the region that is photoheated by the star. The \hii{} region grows in a typical fashion, breaking out of the host halo within $\sim$300 kyr.  The ionization front leaves behind dense neutral clumps that create shadows and form cometary structures similar to ones observed in the Galaxy (also see Figure \ref{fig:projzoom}).  By the end of main sequence, the \hii{} region has grown to 2 proper kpc, enveloped by a partially ionized and heated medium, resulting from the higher energy radiation that has a longer mean free path and can penetrate farther into the neutral IGM.  The shielding from the nearby halos and filaments result in the butterfly shape of the region as seen in previous works \citep{Alvarez2006, Abel2007}. We also show the projections of magnetic energy where $u_{\rm B} = B^2 / 8\pi$ and the square of the fluid vorticity $\omega^2$ where $\mathbf{\omega} = \nabla \times \mathbf{v}$.  The growth of the magnetic field strength is directly related to the vortical fluid motion, and its evolution can be expressed as
\begin{equation}
  \label{eqn:bfield}
  \frac{\partial \mathbf{B}}{\partial t} + \nabla \times (\mathbf{B} \times \mathbf{v}) = 0,
\end{equation}
in the ideal MHD case, i.e. when electrical resistivity is negligible. The vorticity evolution equation can be derived from the Navier-Stokes equation and can be written as
\begin{equation}
  \label{eqn:vort}
  \frac{D \mathbf{\omega}}{Dt} = -\mathbf{\omega} \nabla \cdot \mathbf{v} - 
  \frac{\nabla P \times \nabla \rho}{\rho^2} + \nu \nabla^2 \mathbf{\omega},
\end{equation}
where $\nu$ is the visocity, and we only consider non-viscous fluids ($\nu = 0$) in our simulations.  Here $D/Dt$ is the fluid derivative, $P$ is the pressure, and $\rho$ is the density.  The first term describes the stretching and compression of vortical motions, and the second term comes from non-barotropic flows, $P \neq P(\rho)$, which occur at or near shock fronts.  In the lower panels of Figure \ref{fig:proj}, the presence of vortical structures as shown in the regions of high vorticity imply increased turbulent energy. Because magnetic field amplification is directly related to the vorticity and thus compression, regions of significant magnetic energy and vorticity are co-located with the regions of high density where gravitational collapse has compressed the field lines.

\begin{figure*} 
	\includegraphics[width=\textwidth]{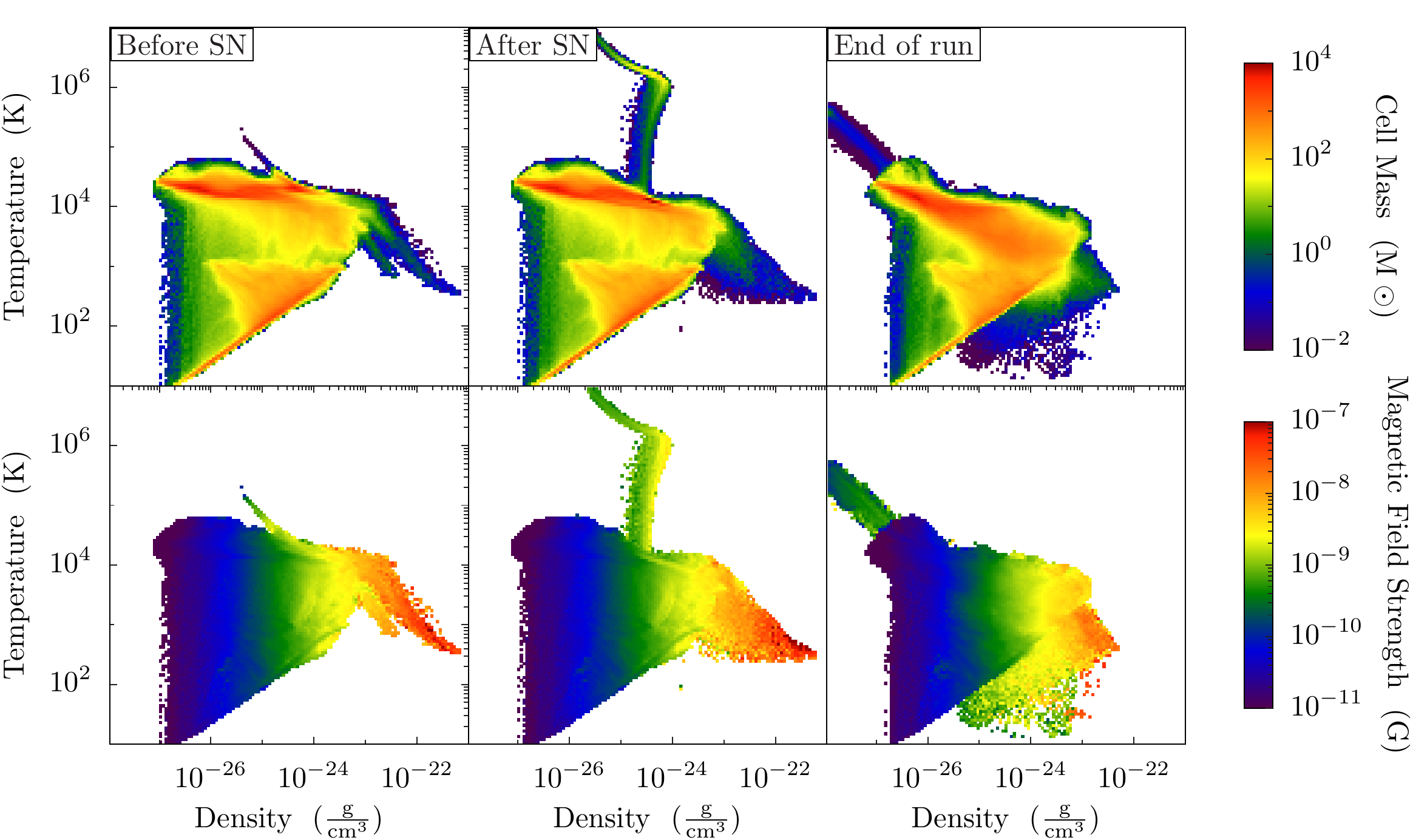}
        \caption{Phase plots of $\rho-T$ right before (left) and after (middle) the supernova, and at the end of the run (right). The top row shows the H2R.B1 run weighted by mass and the bottom row shows the same run weighted by magnetic field strength. The peak, representing the supernova, shows signs of amplification when compared with regions at the same density at lower temperatures.}
    \label{fig:phase}
\end{figure*}

\begin{figure*} 
	\includegraphics[width=\textwidth]{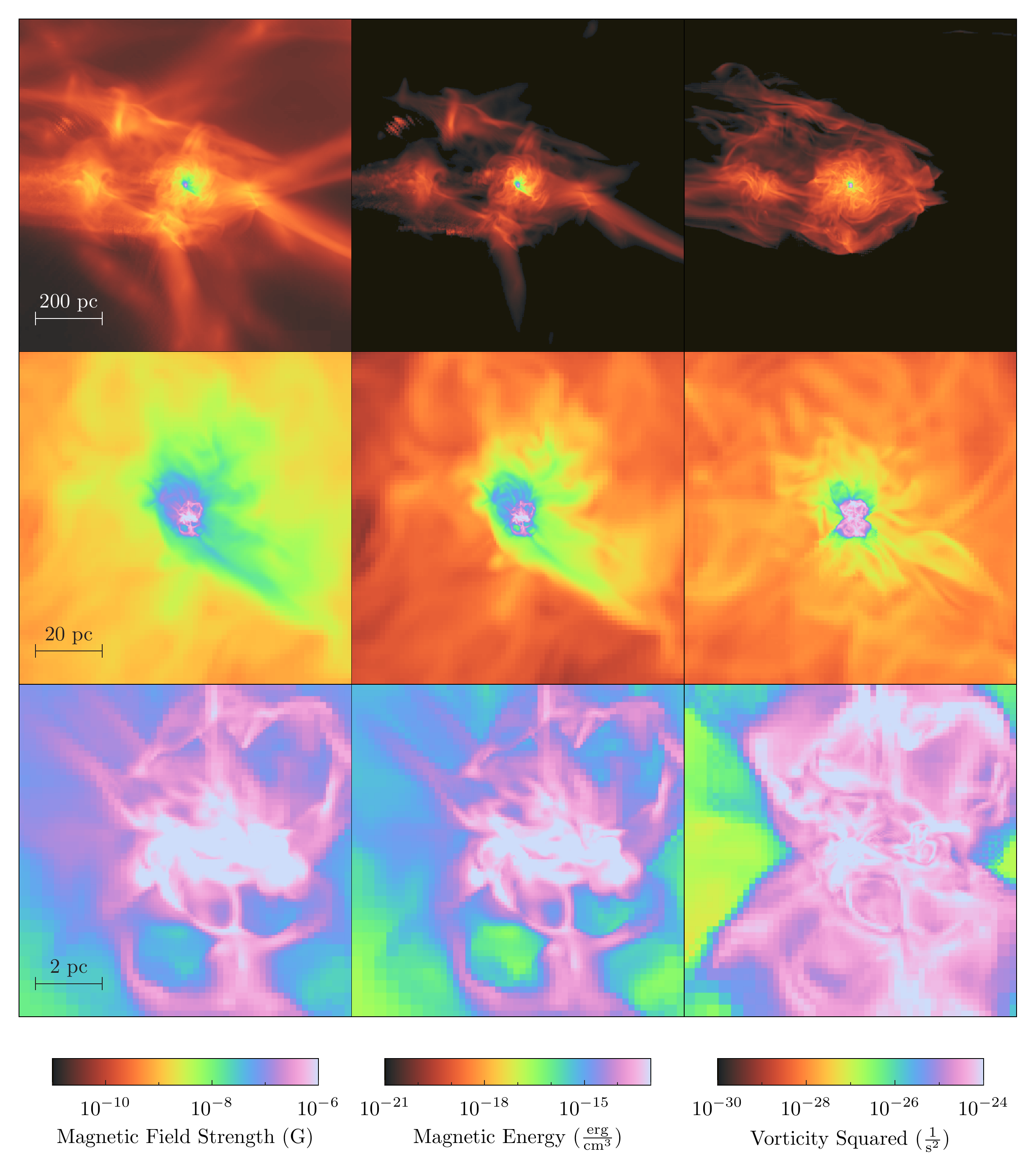}
    \caption{Projections of magnetic field strength, magnetic energy, and vorticity squared centered around the peak vorticity point of the B1 run shortly before the birth of the star. From top to bottom, the widths are 1kpc, 100 pc, and 10 pc. The magnetic fields are highly compressed at this time resulting in amplification. }
    \label{fig:peakvorticity}
\end{figure*}

Figure \ref{fig:projzoom} depicts the same projected quantities of the H2R.B1 run at the birth of the star, the death of the star, and 2 Myr after the supernova with a field of view of 700 pc. In the first row, we see the high density region near the center of the halo where star is formed. This is also the point of peak vorticity in the entire run, arising from the compression of the gas (also see Figure \ref{fig:peakvorticity}). The emitted radiation then evacuates the surrounding gas greatly reducing the baryon density before the death of the star. The ionization fronts also photo-evaporate the gaseous envelopes of some of the nearby halos and filaments, compressing them and producing thin filaments in their shadows. The star lives for about 3.7 million years after which it dies in the form of a Type II supernova. 

In the second row of Figure \ref{fig:projzoom}, there is a clearly delineated shell representing the supernova shock that propagates outwards. The shock mechanically compresses the gas producing regions of enhanced magnetic energy. By the end of the simulations, the supernova has completely disrupted the halo as shown in the bottom row panels. In its wake, the shock leaves behind little knots carrying metals which will eventually dissipate into the ISM. Although the host halo has been completely disrupted, there remains a smaller halo located below the main halo that manages to survive the irradiation and blastwave. This particular halo, now enriched by the metals carried out by the supernova, is likely to be a candidate for hosting second generation star formation. \citet{Smith2015} found that the core of a comparable mini-halo following the supernova of a nearby Pop III star is enriched to $\sim 2 \times 10^{-5}~\mathrm{Z}_{\odot}$.

\subsection{Comparison of radially averaged quantities}

In general, the morphology of the halos is not significantly affected by the presence of the magnetic field.  To make a quantitative comparison, we calculate mass-weighted radial profiles, shown in Figure \ref{fig:profiles}, within a sphere of 1 kpc radius centered on the Pop III star in all three runs at the end of the simulation, about 2 Myr after the supernova. The density and temperature profiles, in particular, show little deviation between the three runs. They also show the approximate location of the supernova shock which at this point is a radius of $\sim 150$ pc. At this point, the remnant is well into the snowplow phase, in line with evolution of the SN remnant as shown in \citet{Greif2007}. The shock has completely blown out the gas reducing the density within the shock radius to $\rho = 10^{-28}$ $\mathrm{g}$ $\mathrm{cm}^{-3}$. Furthermore, the reverse shock heats the gas interior to the remnant initially to $\sim 10^8~\textrm{K}$ and subsequently cools through $PdV$ work to $T = 5 \times 10^{5}$ K. As the shock front expands outwards, the dense shell is able to efficiently cool below $10^4$ K. The temperature and density gradients between shell and the hot interior drive turbulence resulting in magnetic field amplification.

However, notable differences can be seen in the vorticity profile where the difference between the H2R and the H2R.B1 run is more than an order of magnitude inside of the shock radius. For the H2R.B2, the vorticity squared sits between the two runs at $\omega^2 \simeq 6 \times 10^{-29}$ $\mathrm{s}^{-2}$. This difference in vorticity is reflected in the magnetic field strength profiles. The magnetic field strength profile of H2R.B2 has been scaled up by a factor of $10^4$, corresponding to the ratio of initial field strengths, for better comparison with the H2R.B1 run. Recall that the initial seed field strength in the H2R.B1 run was $10^{-10}$ G, 4 orders of magnitude greater than that of H2R.B2. Within the shock radius, the H2R.B2 shows a greater average field strength reflecting the greater vorticity. At this time, the peak magnitudes, which are co-located with the shock radius, are $6.3 \times 10^{-9}$ G for H2R.B1 and $4.2 \times 10^{-13}$ G for H2R.B2. Furthermore, comparing the values shows that the magnitude of the amplification is independent of the initial field strength value because the magnetic field is still dynamically unimportant. 

Figure \ref{fig:phase} shows the $\rho-T$ diagram of the H2R.B1 run immediately before and after the supernova and 2 Myr after the supernova.  The prominent peak in the second column represents the newly formed supernova remnant. As the magnetic field does not affect the dynamics, we do not see any significant differences in the three runs. The bottom row shows the same plot as the top row but shows the mass-weighted average magnetic field strength in each cell rather than the mass. In the bottom middle plot, immediately following the supernova, there is evidence of amplification in the remnant when comparing the field strength at similar densities in the unaffected regions with $T \la 10^4~\textrm{K}$. This peak evolves to lower temperatures as the remnant expands and dissipates into the surrounding medium. The bottom right plot shows the $\rho-T$ diagram at the end of the magnetized run.  The magnetic field within the blastwave and the accompanying shell has been amplified, as seen by the enhanced field strengths below the adiabatic relation in $\rho-T$ phase space and in the hot and diffuse phase.  This additional magnetic energy is not apparent in the bulk of the mass-weighted phase space because of the limited mass affected by the blastwave.

\subsection{Amplification of Magnetic Field}

\subsubsection{Maximum Magnetic Energies}

\begin{figure} 
	\includegraphics[width=\columnwidth]{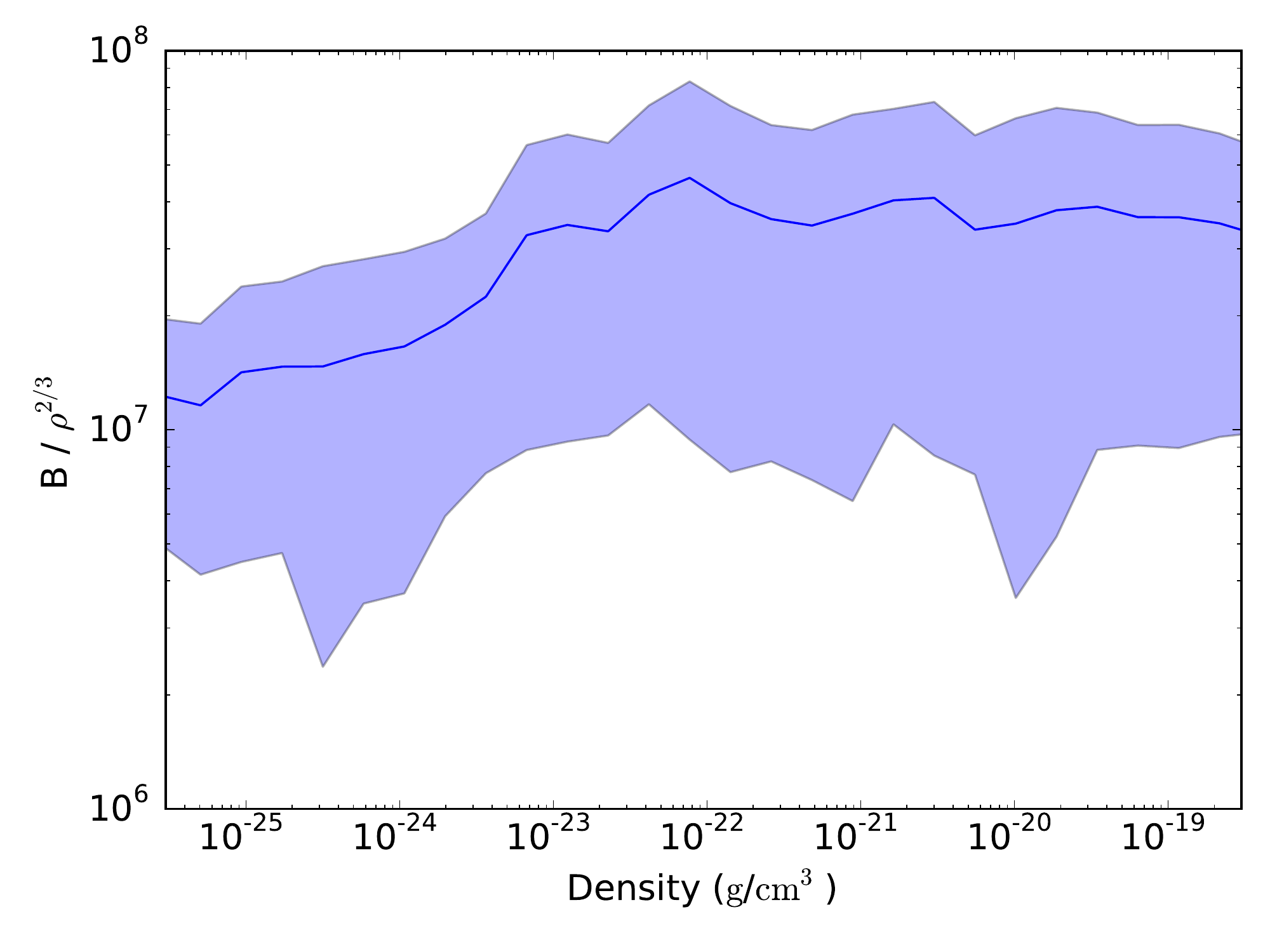}
    \caption{1D mass-weighted profile of the magnetic field strength scaled by $\rho^{2/3}$ against the density at the time of collapse before the formation of the star. The blue line shows the mean where the shaded regions indicate the variance. } 
    \label{fig:bmag}
\end{figure}

To characterize the significance of the magnetic field in this system, we calculated several key values at the point of peak vorticity. Figure \ref{fig:peakvorticity} shows projections of magnetic field and vorticity centered around the point of peak vorticity from the entire simulation. This particular point was found to be at the point of collapse immediately preceding the insertion of the star particle. The peak density at this time is $3.8 \times 10^{-18}~\mathrm{g}$ $\mathrm{cm}^{-3}$ with a magnetic field strength of $3.2 \times 10^{-5}$ G, consistent with the results from \citet{Turk2009} and \citet{Latif2014}. The first is the plasma $\beta \equiv (nkT) / (B^2/8\pi)$ which consistently remains $\beta \gg 1$ throughout the simulation. At the time of collapse, minimum, mean, and max values are 3.0, $4.1 \times 10^6$, and $1.7 \times 10^{11}$, respectively. This implies that the gas dynamics dominate the behavior while magnetic fields have minimal influence. Next, we calculated the Alfv\'{e}nic mach number $M_{\rm A} \equiv V / v_{\rm A}$ where $V$ is the characteristic velocity and $v_{\rm A}$ = $B / \sqrt{4 \pi \rho}$ is the Alfv\'{e}n speed within a sphere of radius $r = 300$ $\mathrm{pc}$ right after the formation of the star. $M_{\rm A}$ remained consistently $M_{\rm A} \gg 1$, typically having values $\sim$1000 outside the shock dropping to $\sim50$ within a pc around the star.  At the time of collapse, minimum, mean, and max values are 2.5, 1400, and $5.4 \times 10^5$. These values also indicate that the magnetic term is not dynamically important.

In Figure \ref{fig:bmag}, we show the magnetic field strength scaled by $\rho^{2/3}$ as a function of density weighted by mass at the time of the halo's collapse. The blue line shows the mean with the shaded region indicating the variance. We see a small deviation from the expected flat relation indicating some dynamo action. Comparing the two relations shows that the field strength to density relation is steeper than 2/3 but not as high as 0.89 as reported by \citet{Turk2009}. This is likely caused by the shorter integration time because we form the star at ~ $1 \times 10^6$ $\rm{cm}^{-3}$, far below $10^{13}$ $\rm{cm}^{-3}$ from \citet{Turk2009}. This is also consistent with the results from \citet{Sur2010} which showed little deviation in the amplification from the $\rho^{2/3}$ relation at a density of $10^{-18}$ $\mathrm{g}$ $\mathrm{cm}^{-3}$.

\subsubsection{Distribution}

In the case of a uniform spherical collapse for a magnetic field frozen into the gas, $B \propto \rho^{\alpha}$ where $\alpha = 2/3$. This relation approximates the amplification due to the compression of magnetic field lines as density increases. Thus, we define the amplification factor to be the ratio 
\begin{equation}
\mathrm{Amplification \ Factor} = \frac{u_{\rm B} \rho^{4/3}}{u_{\rm B_0} \rho_{0}^{4/3}}
\end{equation}
where $u_{{\rm B}_0}$ is the initial seed field energy and $\rho_{0}$ = $\Omega_{\rm b}\rho_c(z = 150)$ is the cosmic mean baryon density. Any value of the amplification factor $> 1$ implies some amplification beyond the compressional scaling which can be attributed to turbulent dynamo effects.

\begin{figure} 
	\includegraphics[width=\columnwidth]{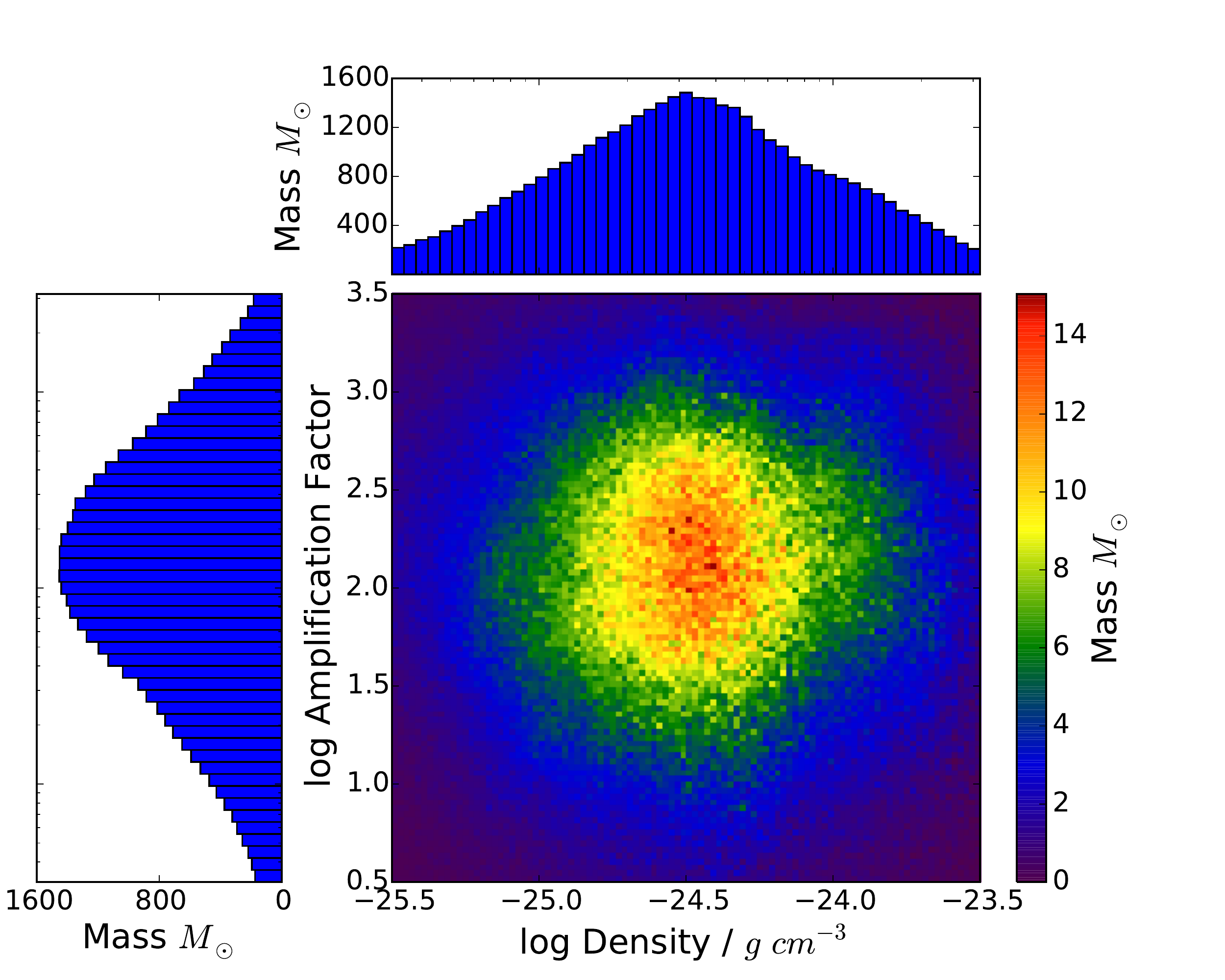}
    \caption{2D mass-weighted histogram of the amplification factor and density at the end of the H2R.B1. Above each respective axes shows the projection to a 1D histogram. The amplification factor shows a clear Gaussian distribution with a mean around 120.}
    \label{fig:hist}
\end{figure}

The phase diagram in Figure \ref{fig:hist} shows this amplification factor as a function of the density weighted by the mass within a sphere of radius 250 pc in run H2R.B1. The blastwave radius is approximately 100 pc at this time. Within this volume, nearly all regions have had its field amplified beyond the expected density scaling, implying dynamo action is efficient during the blast wave propagation, especially during its momentum-conserving phase. To the left of the phase diagram is a histogram showing the distribution of the amplification factor weighted by mass. The amplification factor is log normally distributed with a weighted mean of $10^{2.08}$ and standard deviation of $10^{0.75}$. This is equivalent to a mean field strength amplification by a factor of $\sim$120.

\begin{figure} 
	\includegraphics[width=\columnwidth]{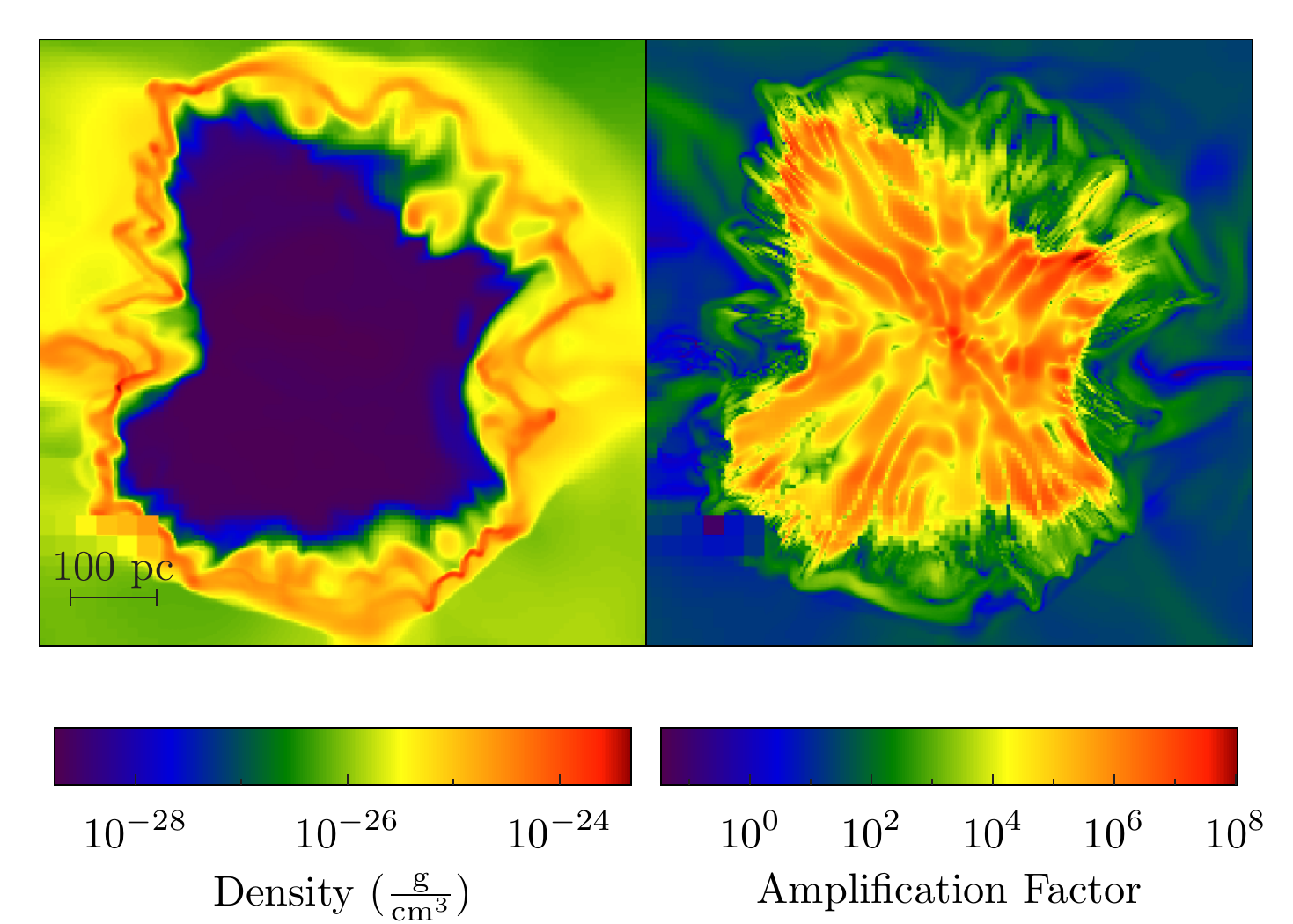}
    \caption{Slice of density (left) and amplification factor (right) centered around the stellar remnant at the end of the H2R.B1 run. Each slice spans 700 pc. Relatively strong magnetic field strengths exist inside the shell having been significantly amplified in the wake of the forward and reverse shocks. The strongest magnetic fields on the order of a few nG exist in the shell.}
    \label{fig:ampfactor}
\end{figure}

Figure \ref{fig:ampfactor} shows a slice of the density and amplification factor at this time showing the distribution of the amplified magnetic field. The relative low densities in the central region evacuated by the supernova leads to high amplification factors. However, the highest magnetic field strengths are located in the shock front, where the gas has been compressed and vortical motions have begun to grow leading to amplification factors on the order of 100.

\subsubsection{Time Evolution}

\begin{figure} 
	\includegraphics[width=\columnwidth]{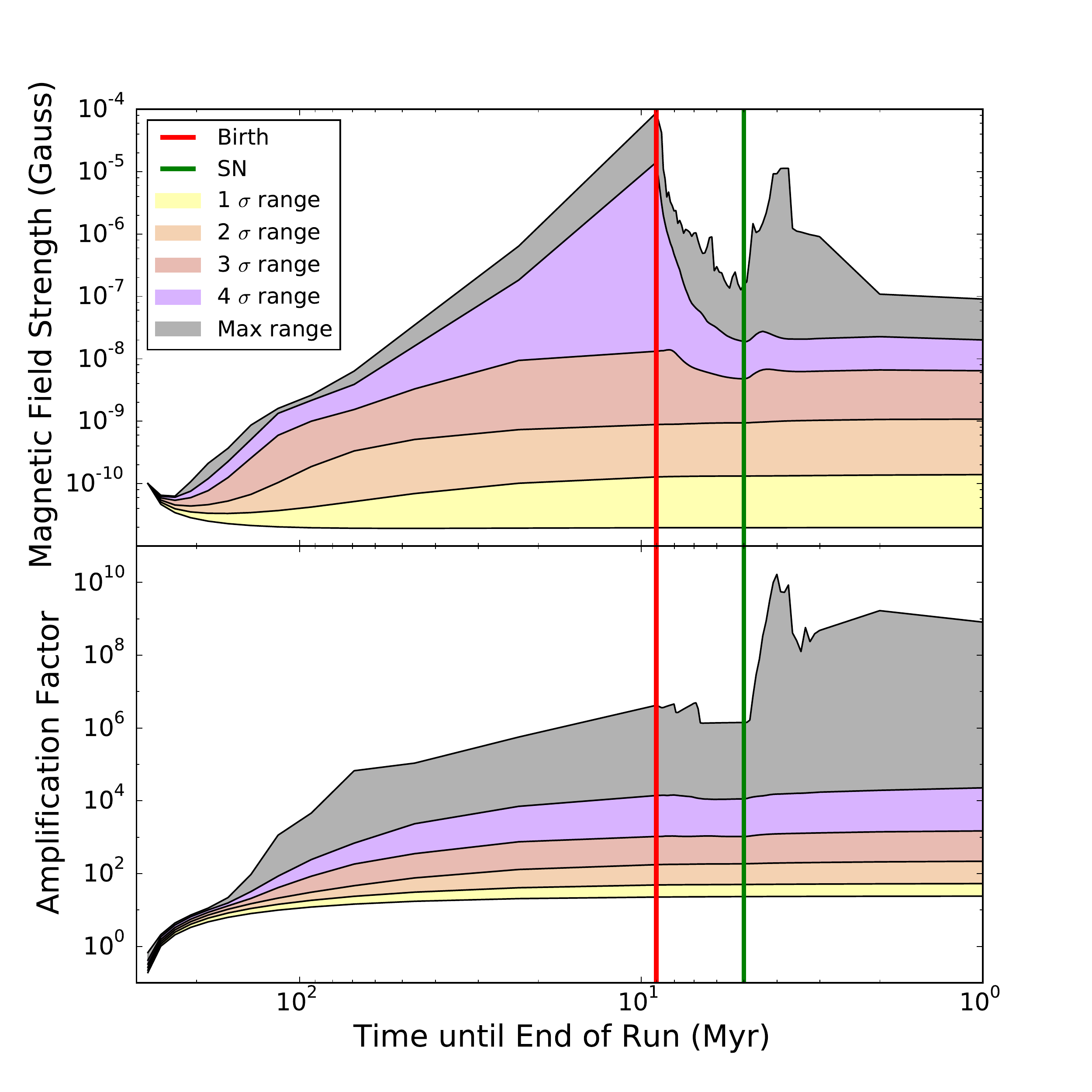}
        \caption{Magnetic field strength (top) and amplification factor (bottom) as a function of time until the end of the simulation in the B1 run. We define the amplification factor to be the ratio $u_{\rm B} / \rho^{4/3}$ normalized by $u_{\rm B_0} / \rho_{0}^{4/3}$ where $u_{\rm B_0}$ is the initial seed field energy and $\rho_{0}$ is the cosmic mean baryon density. $u_{\rm B}$ is defined as $\rm B^2 / 8 \pi$. Only the $+\sigma$ distributions are shown. The field strength shows two peaks, once at the birth of the star and once following the supernova while the amplification factor shows only one peak.  }
    \label{fig:timevol}
\end{figure}

In Figure \ref{fig:timevol}, we show the distribution of the proper field strengths and amplification factor in a sphere of approximately 37.5 kpc centered at the most massive halo which approximately captures the entire Lagrangian volume of the collapsing large-scale environment. The total gas mass in this volume is approximately $3 \times 10^8$ $\mathrm{M}_{\odot}$, which can be used to estimate the gas mass above each multiple of the standard deviation. We plot these quantities as a function of lookback time from the end of the simulation.  Only the top half of the distribution of field strengths is shown through filled in colors while the very bottom line shows the mass-weighted median. The median proper field strength decreases as the scale factor increases since $B \propto a^{2}$. 

In the top panel showing the magnetic field strength, there are two prominent peaks.  The first peak is at the gravitational collapse of the halo immediately prior to the formation of the Pop III star when the density reaches a peak at $3.8 \times 10^{-18}~\mathrm{g}~\mathrm{cm}^{-3}$. As the \hii{} region grows and evacuates the gas from the halo, the magnetic field strength decreases along with the gas density.  The radial forcing and lack of vortical motions in the ionization front suppresses any field amplification.  The second peak follows the death of the star when the supernova produces a shock that compresses the field as it propagates outwards. The first peak hits a maximum at $10^{-4}$ G indicating an amplification of over six orders of magnitude. This is consistent with the results of \citet{Sur2010} who also saw similar levels of amplification. Only a small fraction of the magnetic field in the total volume manages to reach this high level of amplification. While the shock is able to significantly compress the gas, the highest densities are reached at the birth of the star.

The amplification factor evolution differentiates itself from the magnetic field strength evolution with only a single significant peak following the death of the star. To start, the amplification factor shows a sharp increase around 100 Myr before the end of the simulation. This can be attributed to the virialization of the halo generating some turbulence \citep{Wise2007}. Following this period, there is a slight steady increase in the amplification factor as the halo collapses. \citet{King2005} demonstrated that for an anisotropic collapse, $\alpha$ may fluctuate as high as 0.9, where recall $B \propto \rho^{\alpha}$. As the initial peak in the magnetic field evolution was due to compressional effects, which is removed by our scaling of the amplification factor, we observe no significant peak at this point. The most significant amplification occurs following the supernova where the instabilities formed as the supernova cools results in increased turbulence. This turbulent field will then induce stretching and twisting of the magnetic fields through dynamo action resulting in amplification.

In order to get the magnetic field amplification at larger scales, we calculated the magnetic energy spectrum taking a 1 kpc box with a resolution of 0.71 proper pc (AMR level 7) centered about the star particle at the end of the simulation, 2 Myr after the supernova. We found the peak of this spectra to be $k \sim 50 \mathrm{kpc}^{-1}$, corresponding with a coherence length of 20 pc using the definition in \citet{Seifried2014}.  At the end of the run, the radius of the blast wave is around 100 pc. This ratio between the blast wave radius and the coherence length is in agreement with \citeauthor{Seifried2014}.

\begin{figure} 
	\includegraphics[width=\columnwidth]{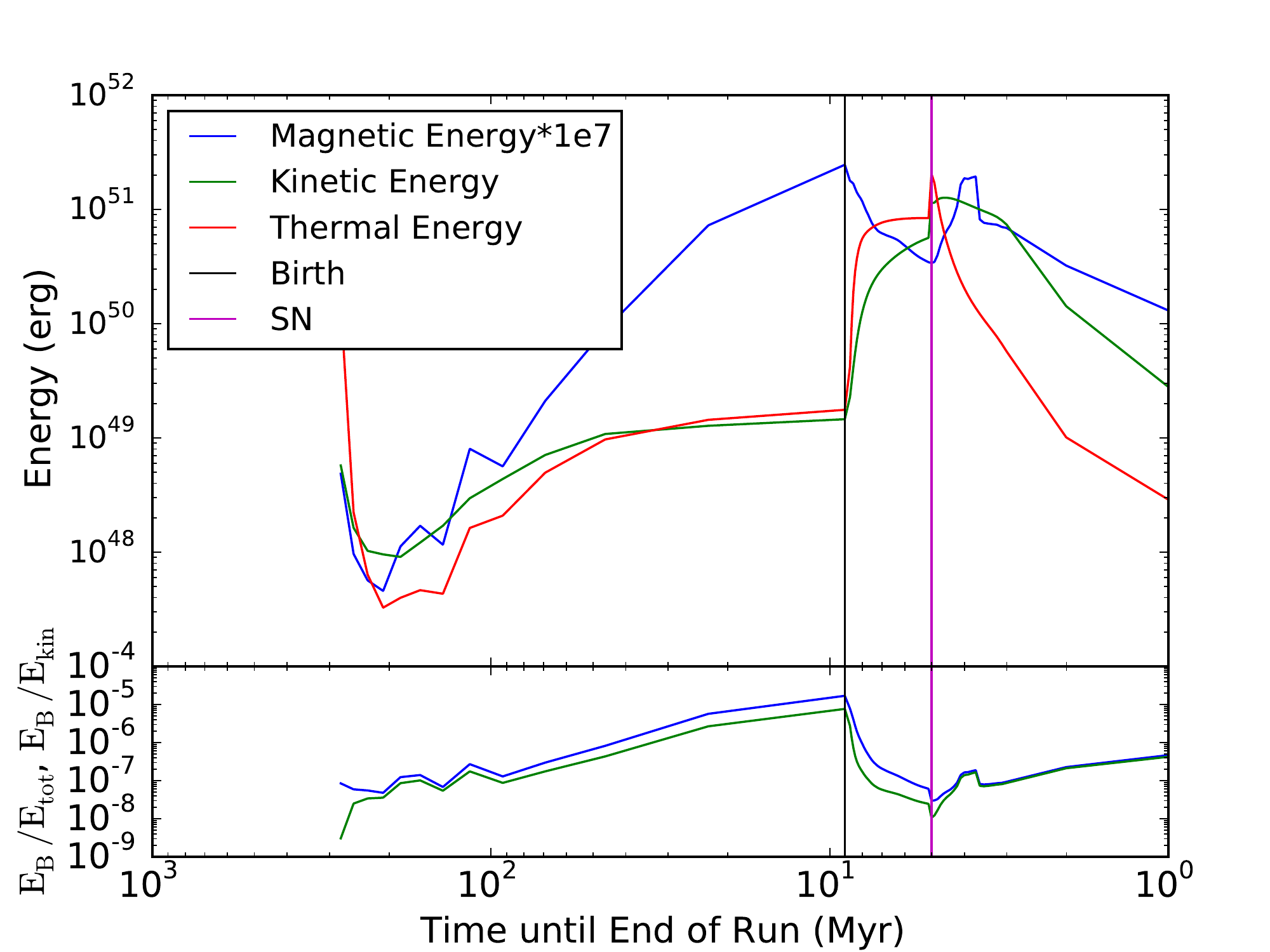}
    \caption{Magnetic, kinetic, and thermal energies as a function of time until the end of the simulation in the B1 run. These total energies are computed within a sphere with a radius of 200 pc, approximately the virial radius of the host halo, centered around the star particle. The magnetic energy has been scaled by $10^7$ for better comparison to the other quantities. The vertical black and magenta lines denote the birth and death of the star.  The panel below shows the ratio of the magnetic energy to the kinetic energy (blue) and total energy (green).}
    \label{fig:energy}
\end{figure}

We also show the time evolution of magnetic, kinetic, and thermal energies in Fig. \ref{fig:energy}. The quantities are the total energies within a sphere of radius 200 pc, which is approximately the virial radius of the host halo, centered around the star particle. The evolution of the magnetic energy shows the two peaks previously described in Fig. \ref{fig:timevol}. In the bottom panel, we plot the ratio of the magnetic energy to both the kinetic energy and total energy. At the time of collapse, when the magnetic energy is at a global maximum, we see that the kinetic energy dominates the magnetic term by 5 orders of magnitude. This shows that the magnetic term is never dynamically significant consistent with our earlier conclusions.

\section{Discussion}
\label{sec:disc}

Our simulations show that magnetic fields are amplified mostly strongly via self-consistent turbulence generated by mechanical compression and the initial field strength plays little role in the subsequent level of amplification.


In our simulations, we have required that the Jeans length be resolved by 64 cells along each dimension. As \citet{Turk2012} and \citet{Sur2010} has shown, although 64 cells is sufficient to resolve the action of the dynamos, it may not be enough to fully resolve the amplification of the fields as a result of the dynamos. Our simulations do not show any signs of magnetic saturation, and thus equipartition, and our results can thus be taken as a lower limit to the field strength.  However, due to computational constraints, we were unable to increase the resolution preventing any declarative statement about convergence.

In our simulations, we have only considered the ideal MHD limit in which the flux-freezing approximation holds. The only dissipation observed is a numerical artifact resulting from the finite resolution of the simulation. We do not consider the effects of Ohmic resistivity, ambipolar diffusion, nor magnetic reconnection. Although the calculated values of plasma $\beta \gg 1$ and $M_{\rm A} \gg 1$ validate the approximation, our need for higher resolution may require taking non-ideal effects into account as a result. In particular, the effects due to magnetic reconnection in the vicinity of the star may produce significant deviations in the amplification process.

Missing physics that may have dynamical effects include streaming velocities and stellar magnetic fields. First, the relative velocity differences of dark matter and baryonic gas \citep{Tseliakhovich2011} results in a delayed collapse of halos which may have important dynamical impact \citep[e.g.][]{Greif2011, Stacy2011, OLeary2012}. Namely, the increased velocity in the gas may produce a greater shearing effect which would increase the turbulence, invoking greater amplification of the magnetic fields.

Secondly, stars can themselves generate powerful magnetic fields. A fraction of galactic O-type stars with masses up to 60 $\mathrm{M}_{\odot}$  have been observed to have surface magnetic field strengths of $\sim 100$ G \citep{Wade2014}. Moreover, magnetic fields in protostellar disks can be sufficiently amplified leading to field strengths capable of driving jets \citep{Latif2016}. These fields are coherent at scales up to 1000 AU with a corresponding jet luminosity of $\sim 10^6$ $\mathrm{L}_{\odot}$. Furthermore, the magnetic fields produced in the form of supernova feedback can also play a significant role \citep{Schober2013}. While these fields may be significantly below our current effective computational resolution, future simulations where higher resolutions are demanded may need to include these effects.

\section{Conclusions}
\label{sec:conc}

In this paper, we present the amplification of the magnetic field in the \hii{} region throughout the lifetime of a single Pop III star and its supernova. We simulated three different runs including a base case without any magnetic fields, and two others with an uniform initial background proper field strength of $10^{-10}$ G and $10^{-14}$ G. In each simulation, a single Pop III star of 40 $\mathrm{M}_{\odot}$ forms in the most massive halo at $z \sim$ 15 in the central metal-free molecular cloud and subsequently emits radiation until its death in the form of a supernova injecting $10^{51}$ erg into its surroundings. The simulation ends after about 2 Myr after the death of the star as the shockwave continues to propagate outwards. We tracked the evolution of the magnetic field throughout each of the simulations and found the following main results.
\begin{enumerate}
\item Magnetic fields are amplified primarily through compression during the gravitational collapse prior to star formation and scales as $\rho^{2/3}$ as expected from ideal collapse scenarios.
\item We find no significant amplification during the growth of the \hii{} region as the star evacuates the gas from its host halo and photoevaporates nearby halos and filaments.
\item Once the supernova remnant begins to cool and fragment, the resulting turbulent velocity in and near the the supernova shell further amplify the magnetic field through small-scale dynamo action. Here the field strengths have a log-normal distribution with an average amplification factor of 120. Within the shell, the field strength is on the order of a few nG at a number of 1 $\rm{cm}^{-3}$.
\item The amplitude of the amplification is largely independent of the initial seed field strength. The peak level of amplification occurs in the interior of the blastwave, where the resulting field strength is six orders of magnitude greater than the amplification levels expected in a spherical collapse.
\end{enumerate}

Our simulations show the potential for dynamically important magnetic fields to be produced in the first galaxies. With stronger background field strengths closer to observed limits and the inclusion of fields generated by stars, the amplification mechanisms described in this paper can produce dynamically important fields. Our work elucidates the magnetic field ``initial conditions'' in the protogalactic gas that will collapse in descendant halos, forming low-mass metal-enriched galaxies.  Future calculations will follow its evolution to study the impact of magnetic fields on the transition from Population III stars to the first generations of galaxies, possibly affecting the nature of star formation in such objects.

\section*{Acknowledgements}

We like to thank our referee, Robi Banerjee, who provided helpful comments
to strengthen our work. This research was supported by National Science Foundation (NSF)
grants AST-1211626 and AST-1333360 and Hubble Theory grants
HST-AR-13895 and HST-AR-14326.  Support for programs \#13895 and
\#14326 were provided by NASA through a grant from the Space Telescope
Science Institute, which is operated by the Association of
Universities for Research in Astronomy, Inc., under NASA contract NAS
5-26555.  The simulations were performed on Comet operated by the San
Diego Supercomputer Center with the XSEDE allocation AST-120046.  This
work was performed using the open-source {\sc Enzo} and {\sc yt}
codes, which are the products of collaborative efforts of many
independent scientists from institutions around the world. Their
commitment to open science has helped make this work possible.




\bibliographystyle{mnras}
\bibliography{library} 




%


\bsp	
\label{lastpage}
\end{document}

%% file: macros.tex
\newcommand{\fesc}{\ifmmode{f_{\rm esc}}\else{$f_{\rm esc}$}\fi}
\newcommand{\fescs}{\ifmmode{f_{\rm esc}^\star}\else{$f_{\rm esc}^\star$}\fi}
\newcommand{\kms}{\ifmmode{{\;\rm km~s^{-1}}}\else{km~s$^{-1}$}\fi}
\newcommand{\fgas}{\ifmmode{{f_{\rm gas}}}\else{$f_{\rm gas}$}\fi}
\newcommand{\cubecm}{\ifmmode{{\rm cm^{-3}}}\else{cm$^{-3}$}\fi}
\newcommand{\ztwo}{\ifmmode{{\rm [Z_2/H]}}\else{[Z$_2$/H]}\fi}
\newcommand{\zthree}{\ifmmode{{\rm [Z_3/H]}}\else{[Z$_3$/H]}\fi}
\newcommand{\lsim}{\lower0.3em\hbox{$\,\buildrel <\over\sim\,$}}
\newcommand{\gsim}{\lower0.3em\hbox{$\,\buildrel >\over\sim\,$}}

\newcommand{\sfr}{\ifmmode{\textrm{M}_\odot \,\textrm{yr}^{-1} \,\textrm{Mpc}^{-3}}\else{M$_\odot$ yr$^{-1}$ Mpc$^{-3}$}\fi}
\newcommand{\hsfr}{\ifmmode{\textrm{M}_\odot\, \textrm{yr}^{-1}}\else{M$_\odot$ yr$^{-1}$}\fi}

\newcommand{\eavg}{\ifmmode{\langle E_\gamma \rangle}\else{$\langle E_\gamma \rangle$}\fi}

\newcommand{\enzo}{{\sc enzo}}

\newcommand{\Ms}{\ifmmode{M_\odot}\else{$M_\odot$}\fi}
\newcommand{\vrms}{\ifmmode{v_{\rm rms}}\else{$v_{\rm rms}$}\fi}

\newcommand{\tvir}{\ifmmode{T_{\rm{vir}}}\else{$T_{\rm{vir}}$}\fi}
\newcommand{\mvir}{\ifmmode{M_{\rm{vir}}}\else{$M_{\rm{vir}}$}\fi}
\newcommand{\rvir}{\ifmmode{r_{\rm{vir}}}\else{$r_{\rm{vir}}$}\fi}

\newcommand{\jj}{\ifmmode{J_{21}}\else{$J_{21}$}\fi}
\newcommand{\flw}{\ifmmode{F_{LW}}\else{$F_{LW}$}\fi}
\newcommand{\kph}{\ifmmode{k_{\rm ph}}\else{$k_{\rm ph}$}\fi}

\newcommand{\zsun}{\ifmmode{\rm\,Z_\odot}\else{$\rm\,Z_\odot$}\fi}

\newcommand{\hi}{H {\sc i}}
\newcommand{\hii}{H {\sc ii}}
\newcommand{\hei}{He {\sc i}}
\newcommand{\heii}{He {\sc ii}}
\newcommand{\heiii}{He {\sc iii}}
\newcommand{\nhi}{\ifmmode{N_{\rm HI}}\else{$N_{\rm HI}$}\fi}
\newcommand\unit[1]{\; \textrm{#1}}

\def\eps@scaling{1.0}%
\newcommand\epsscale[1]{\gdef\eps@scaling{#1}}%
\newcommand\plotone[1]{%
 \centering 
 \leavevmode 
 \includegraphics[width={\eps@scaling\columnwidth}]{#1}%
}%
\newcommand\plottwo[2]{%
 \centering 
 \includegraphics[width={\eps@scaling\columnwidth}]{#1}%
 \hfil 
 \includegraphics[width={\eps@scaling\columnwidth}]{#2}%
}%